\documentclass[11pt,showpacs]{revtex4}
\input epsf.sty

\usepackage{ulem}
\usepackage{cancel}
\usepackage{color}

\def\comment#1{}

\usepackage{graphicx}
\begin{document}
\title{Why is the top quark much heavier than other fermions?}
\author{She-Sheng Xue}
\email{xue@icra.it}
\affiliation{ICRANeT, Piazzale della Repubblica, 10-65122, Pescara,\\
Physics Department, University of Rome ``La Sapienza'', Rome,
Italy} 
%\affiliation{$^{(b)}$Department of Physics, Isfahan University of Technology, Isfahan 84156-83111, Iran}

%\date{Received version \today}

\begin{abstract}
The recent ATLAS and CMS experiments show the first observations of a new particle in the search for the Standard Model Higgs boson at the LHC. We revisit the theoretical inconsistency of the fundamental high-energy cutoff with the parity-violating gauge symmetry of local quantum field theory for the standard model. This inconsistency suggests high-dimensional operators of fermion interactions, which are attributed to the quantum gravity. In this Letter, 
recalling the minimal dynamical symmetry-breaking mechanism, we show that it is energetically favorable for the top quark to acquire its mass via spontaneous symmetry breaking, whereas other fermions acquire their masses via explicit symmetry breaking.    
\end{abstract}

\pacs{12.60.-i,12.60.Rc,11.30.Qc,11.30.Rd,12.15.Ff}

\maketitle

%\vskip0.1cm
\noindent
{\bf Introduction.}
\hskip0.1cm
Since its appearance, the Standard Model for elementary particle physics has always been extremely peculiar. The parity-violating gauge couplings, the hierarchy of fermion masses, and flavor mixing have been at the center of a conceptual elaboration and
an intensive experimental analysis that have played a major role in
donating to mankind the beauty of the Standard Model (SM) for particle physics. Chiral gauge symmetries on the one hand and spontaneous/explicit breakings of these symmetries on the other play essential roles in understanding the parity-violating gauge couplings and mass spectra of fermions in the Standard Model. The Nambu-Jona-Lasinio model \cite{njl} for high energies and its effective counterpart for low energies, the Higgs model \cite{higgs}, provides an elegant description for the
electroweak-breaking-scale, intermediate gauge boson masses and their relations. Nevertheless, we are still searching for a physical mechanism to explain the hierarchy of the fermion masses and the flavor mixing angles. Much theoretical effort has been made on this central issue since the Standard Model was established. Likewise, after a great experimental effort for many years, the ATLAS \cite{ATLAS} and CMS \cite{CMS} experiments have recently shown the first observations of a new particle in the search for the standard
model Higgs boson at the LHC. This far-reaching result
begins to shed light on this most elusive and fascinating arena of fundamental particle physics. 

Assuming {\it the dynamical symmetry breaking} to dynamically generate the electroweak scale is theoretically rather attractive. Much effort has been made in this direction since the Standard Model was born. Apart from their theoretical problems, such as the fine-tuning and hierarchy problems, the models of the dynamical symmetry breaking are still far from quantitatively explaining experimental results, particularly given the recent discovery of the 125 GeV scalar particle \cite{ATLAS,CMS}. Nevertheless, this newly discovered particle spurs it is a impetus and worthwhile for us to revisit the theoretical issue of the dynamical symmetry breaking to find any dynamical or symmetrical aspect that requires further studies to satisfactorily explain the experimental results. 

When the top quark mass $m_t$ was discovered to be greater than $\sim 10^2\,$GeV, several authors \cite{nambu1989,Marciano1989,mty1989,bhl1990} in 1989 suggested that the symmetry breakdown of the Standard Model could be a dynamical mechanism of the Nambu-Jona-Lasinio or BCS type that intimately involves the top quark at a high-energy scale $\Lambda$. This dynamical mechanism leads to the formation of a low-energy $\bar t t$-condensate, which is responsible for the top quark, $W^\pm$ and $Z^\circ$ gauge bosons masses, and a composite particle of the Higgs type.
Since then, many models based on this idea have been proposed and studied \cite{DSB_review}. 
For our discussions on this idea, we will adopt the model for the minimal dynamical symmetry breaking via an effective four-fermion operator of the Nambu-Jona-Lasinio type
\begin{eqnarray}
L = L_{\rm kinetic} + G(\bar\Psi^{ia}_Lt_{Ra})(\bar t^b_{R}\Psi_{Lib}),
\label{bhl}
\end{eqnarray}
which was studied by Bardeen, Hill and Lindner (BHL) \cite{bhl1990} in the context of a well-defined quantum field theory at the high-energy scale $\Lambda$; the coupling $G$ is on the order of $1/\Lambda^2$.

To achieve the low-energy electroweak scale for the top quark mass $m_t$ by the renormalization group equations \cite{bhl1990,Marciano1989,bhl1990a}, this model (\ref{bhl}) requires $\Lambda/m_t \gg 1$ with a drastically unnatural fine tuning, which is known as the gauge hierarchy problem, and the top quark mass $m_t$ is determined by the infra red quasi-fixed point \cite{bhl1990a}. To have a natural scheme incorporating the effective four-fermion operator of the Nambu-Jona-Lasinio type (\ref{bhl}), some strong technicolor (TC) dynamics at the $\sim$ TeV scale were invoked \cite{hill1994}; this scheme is preferentially coupled to the third quark family of top and bottom quarks. In addition, from the phenomenological point view, the newly discovered 125 GeV particle does not seems to be the neutral $\bar t t$-composite scalar that is significantly heavier than the top quark mass \cite{bhl1990}. The possibility of the $125$ GeV particle being a light pseudoscalar, such as the top-pion \cite{bhl1990a}, seems unlikely because the loop-suppressed couplings of light pseudoscalars to the SM gauge bosons are too small to generate the observed signal \cite{top-pion}.

These discussions indicate that much effort is still required to study the issue of the minimal dynamical symmetry breaking that preferentially associated with the top quark (the top-Higgs system) in the theoretical aspects of dynamics or/and symmetry (see for example \cite{Hill2013}) to discover if the issue agree with experiments.  
In this Letter, we will focus on the theoretical question of why the dynamical symmetry breaking is minimally or preferentially associated with the top quark. We assume that the high-dimensional operators of all fermion fields could be attributed to the new dynamics, such as the quantum gravity at the Planck scale $\Lambda_{\rm pl}$. For example, the four-fermion operator in the Einstein-Cartan theory can be obtained by integrating over static torsion fields at the Planck scale. It is conceivable that the new dynamics at the scale $\Lambda$ should be on an equal footing with all the fermions in the Standard Model because the scale $\Lambda$ is much larger than the masses of all the fermions. This finding raises a neutral question: why should the new dynamics preferentially act on the top quark alone? It is the aim of this Letter to understand, from the dynamical point of view, a compelling possible answer to this question.
Within the context of the Standard Model and dynamical symmetry breaking, we attempt to show that the minimal dynamical symmetry breaking (\ref{bhl}) for the top quark, by which this particle acquires its dynamical mass, is an energetically favorable configuration (the ground state) of the quantum field theory with the high-dimension operators of all the fermion fields at the cutoff $\Lambda$. 

To explain why the top quark is much heavier than the other fermions, some discussions of the origins of fermion masses are required; the top quark mass is attributed to the dynamical symmetry breaking, whereas the other fermion masses are attributed to explicit symmetry breakings. In addition to the broken phase where the dynamical symmetry breaking occurs, we will discuss the symmetric phase for strong couplings where the dynamics of high-dimension operators of fermion fields form the massive composite states of three fermions the preserve the chiral gauge symmetries of the Standard Model. This dynamical feature gives a possibility to solve the fine-tuning problem and may hint at the composite scalar mass. 
The natural units $\hbar=c=1$ are adopted, unless otherwise specified.

%\vskip0.1cm
\noindent
{\bf Dynamical symmetry breaking of the third quark-family.}
\hskip0.1cm 
To simplify the discussions and calculations, we first consider the third quark family only, the left-handed doublet $\Psi_L=(t_L,b_L)$ and the right-handed singlet $\psi_R=t_R,b_R$, and generalize the BHL proposal (\ref{bhl}) as follows: 
\begin{eqnarray}
L &=& L_{\rm kinetic} + G(\bar\Psi^{ia}_L\psi_{Rja})(\bar \psi^{jb}_R\Psi_{Lib}),\nonumber\\
&=& L_{\rm kinetic} + G(\bar\Psi^{ia}_Lt_{Ra})(\bar t^b_{R}\Psi_{Lib})
+ G(\bar\Psi^{ia}_Lb_{Ra})(\bar b^b_{R}\Psi_{Lib}),
\label{bhlx}
\end{eqnarray}
where $a,b$ and $i,j$ are, respectively, the color and flavor indexes of the top and bottom quarks. The fermion fields in $L_{\rm kinetic}$ are supposed to be massless. This Lagrangian has not only an $SU_c(3)\times SU_L(2)\times U_Y(1)$ gauge symmetry of the Standard Model but also a global $SU_L(2)\times U_R(1)$ flavor symmetry.

Following the BHL calculations, we have the gap equations for the induced top- and bottom-quark masses $m_t=-G\langle \bar tt\rangle$ and $m_b=-G\langle \bar bb\rangle$:
\begin{eqnarray}
\left(\matrix{m ~~0\cr ~~~0 ~~~ m} \right) &=& 2GN_c\frac{i}{(2\pi)^4}\int d^4l(l^2-m^2)^{-1}\left(\matrix{m ~~0\cr ~~~0 ~~~ m} \right),
\label{gap0}
\end{eqnarray}
where $m=m_t=m_b$. The result of evaluating Eq.~(\ref{gap0}) with a momentum-space cutoff $\Lambda$ is    
\begin{eqnarray}
G^{-1} &=& \frac{N_c}{8\pi^2}[\Lambda^2-m^2\ln (\Lambda^2/m^2)].
\label{gap1}
\end{eqnarray}
In addition to the trivial solution $m=0$, the gap equation (\ref{gap0}) has a nontrivial solution $m\not=0$ for a sufficiently strong coupling, $G\geq G_c=8\pi^2/(N_c\Lambda^2)$, where $G_c$ is the ``critical'' coupling constant. The nontrivial solution $m_t=m_b=m\not=0$ to the gap equation (\ref{gap1}) is valid for both the $t$-channel and the $b$-channel. The gap equation (\ref{gap1}) can be written as 
\begin{eqnarray}
\frac{1}{G_c}-\frac{1}{G}=(1/G_c)(m/\Lambda)^2\ln (\Lambda/m)^2= (N_cm^2/8\pi^2)\ln (\Lambda/m)^2>0.
\label{delta}
\end{eqnarray}
When $m\ll \Lambda$, the four-fermion coupling $G=G(\Lambda, m_t)\rightarrow G_c$.

As will be shown below, the energetically favorable configuration of this defined 
quantum field theory (\ref{bhlx}) should be the configuration ($m_t=m\not=0, m_b=0$), rather than the configuration ($m_t=m_b=m\not=0$). %and ($m_t=m_b=m=0$).     
To calculate all possible bubble diagrams contributing to the vacuum energy of such defined 
quantum field theory (\ref{bhlx}), we must identify all the elementary and composite modes when the gap equation (\ref{gap1}) is satisfied for $G\gtrsim G_c$ as well as the two-point Green functions of these modes. 

Following the BHL calculations \cite{bhl1990}, we use the four-fermion interacting vertexes $G(\bar\Psi^{ia}_Lt_{Ra})(\bar t^b_{R}\Psi_{Lib})
$ ($t$-channel) and $G(\bar\Psi^{ia}_Lb_{Ra})(\bar b^b_{R}\Psi_{Lib})$ ($b$-channel) in the effective Lagrangian (\ref{bhlx}) to calculate the gap equation (\ref{gap0}) or (\ref{gap1}) for $m_t=m_b=m\not=0$. In addition, 
we obtain the following:
(1)
the inverse propagators of the top $t$ and bottom $b$ quarks:
\begin{eqnarray}
\Gamma^{-1}_{t,b}(p^2, m_{t,b}) &=& (\gamma_\mu p^\mu-m_{t,b});
\label{fermion}
\end{eqnarray}
(2) two composite scalar modes for the $t$- and $b$-channels and the inverse propagators of these modes
\begin{eqnarray}
\Gamma^{-1}_S(p^2, m_{t,b}) &=& 2N_c (p^2-m_{t,b}^2)(4\pi)^{-2}\int_0^1dx \ln \left\{\Lambda^2/[m^2_{t,b}-x(1-x)p^2]\right\};
\label{scalar}
\end{eqnarray}
(3) two composite neutral-pseudoscalar modes for the $t$- and $b$-channels and the inverse propagators of these modes:
\begin{eqnarray}
\Gamma^{-1}_P(p^2, m_{t,b}) &=& 2N_c p^2(4\pi)^{-2}\int_0^1dx \ln \left\{\Lambda^2/[m^2_{t,b}-x(1-x)p^2]\right\};
\label{pscalar}
\end{eqnarray}
(4) two composite charged-pseudoscalar modes for the $t$- and $b$-channels and the inverse propagators of these modes
\begin{eqnarray}
\Gamma^{-1}_F(p^2, m_{t,b}) &=& 8N_c p^2(4\pi)^{-2}\int_0^1dx (1-x) \ln \left\{\Lambda^2/[m^2_{b,t}x+m^2_{t,b}(1-x)-x(1-x)p^2]\right\}.
\label{cpscalar}
\end{eqnarray}
In addition to these composite modes, there are elementary modes of the gauge bosons $W_\mu^\pm$, $Z_\mu^\circ$ and photon $\gamma$. The inverse propagators of these modes are
\begin{eqnarray}
\Gamma^{-1}_{\mu\nu}(p^2, M^2_{W,Z,\gamma}) &=& (p_\mu p_\nu/p^2-g_{\mu\nu})Z_{W,Z,\gamma}(p^2)(p^2-M^2_{W,Z,\gamma}),
\label{boson}
\end{eqnarray}
where the masses $M^2_{W}=M^2_{Z}\sin^2\theta_W$, $M^2_{\gamma}=0$ and the wave-function renormalization constants $Z_{W,Z,\gamma}(p^2)$. One of two composite neutral-pseudoscalar modes (\ref{pscalar}) becomes the longitudinal component of the intermediate gauge boson $Z_\mu^\circ$. One of two composite charged-pseudoscalar modes (\ref{cpscalar}) becomes the longitudinal component of the intermediate gauge boson $W_\mu^\pm$. We assume these composite modes come from the $t$-channel.
As a result, the composite modes that remain are two massive scalar modes of the Higgs type from the $t$-channel and $b$-channel as well as one neutral-pseudoscalar mode and one charged-pseudoscalar mode from the $b$-channel.    

We turn to the configuration ($m_t=m\not=0$) and ($m_b=0$). Using the interacting vertex $G(\bar\Psi^{ia}_Lb_{Ra})(\bar b^b_{R}\Psi_{Lib})$ in Eq.~(\ref{bhlx}) for the $b$-channel and $m_b=0$, we calculate the inverse two-point 
Green functions analogously to Eqs.~(\ref{scalar}-\ref{boson}) and obtain
\begin{eqnarray}
\Gamma^{-1}_S(p^2) &=& (2/G)+2N_c p^2(4\pi)^{-2}\int_0^1dx \ln \left\{\Lambda^2/[-x(1-x)p^2]\right\},
\label{scalarb}\\
\Gamma^{-1}_P(p^2) &=& (2/G)+ 2N_c p^2(4\pi)^{-2}\int_0^1dx \ln \left\{\Lambda^2/[-x(1-x)p^2]\right\},
\label{pscalarb}\\
\Gamma^{-1}_F(p^2) &=& (2/G) + 8N_c p^2(4\pi)^{-2}\int_0^1dx (1-x) \ln \left\{\Lambda^2/[m^2_t(1-x)-x(1-x)p^2]\right\}.
\label{cpscalarb}
\end{eqnarray}
Unlike Eqs.~(\ref{scalar}-\ref{cpscalar}), the first term $(2/G)$ remains due to the absence of the gap equation for $m_b\not=0$ in the $b$-channel. As a result,
the inverse two-point 
Green functions (\ref{scalarb}-\ref{cpscalarb}) do not represent inverse propagators of massive or massless composite modes. 
In Eqs.~(\ref{fermion},\ref{boson}) and Eqs.~(\ref{scalarb}-\ref{cpscalarb}),
we are allowed to introduce the $b$-quark mass $\bar m_b\not=0$ $(\bar m_b\ll m_t$) by a small explicit breaking of the chiral symmetries in the kinetic term $L_{\rm kinetic}$ (\ref{bhlx}). In this case, Eqs.~(\ref{scalarb}-\ref{cpscalarb}) are replaced by
\begin{eqnarray}
\Gamma^{-1}_S(p^2) &=& 2\delta +2N_c (p^2-4\bar m_b^2)(4\pi)^{-2}\int_0^1dx \ln \left\{\Lambda^2/[\bar m_b^2-x(1-x)p^2]\right\},
\label{scalarbb}\\
\Gamma^{-1}_P(p^2) &=& 2\delta+ 2N_c p^2(4\pi)^{-2}\int_0^1dx \ln \left\{\Lambda^2/[\bar m_b^2-x(1-x)p^2]\right\},
\label{pscalarbb}\\
\Gamma^{-1}_F(p^2) &=& 2\delta + 8N_c p^2(4\pi)^{-2}\int_0^1dx (1-x) \ln \left\{\Lambda^2/[\bar m_b^2 x + m^2_t(1-x)-x(1-x)p^2]\right\},
\label{cpscalarbb}
\end{eqnarray}
and
\begin{eqnarray}
\delta &\equiv& \frac{1}{G}-\frac{1}{G_c}\left(1-(\bar m_b/\Lambda)^2\ln (\Lambda/\bar m_b)^2\right)\label{delta0}\\
&=&\frac{1}{G_c}\left(\frac{\bar m_b}{\Lambda}\right)^2\left\{\left[ \left(\frac{m_t}{\bar m_b}\right)^2-1\right]\ln \left(\frac{\Lambda}{\bar m_b}\right)^2 -\left(\frac{m_t}{\bar m_b}\right)^2\ln \left(\frac{m_t}{\bar m_b}\right)^2\right\}\nonumber\\
&\approx &\frac{1}{G_c}\left(\frac{ m_t}{\Lambda}\right)^2 \ln \left(\frac{\Lambda}{m_t}\right)^2 > 0,\quad m_t\gg \bar m_b
\label{delta1}
\end{eqnarray}
where the $\delta$ term contains the contribution $1/G$ of the four-fermion vertex and the contribution [the second part in line (\ref{delta0})] of 
two disconnect tadpole diagrams in the $b$-channel, and we use the gap equation (\ref{delta}) for the top-quark mass $m_t$
to obtain Eq.~(\ref{delta1}).

%\vskip0.1cm
\noindent
{\bf The vacuum energy of quantum field theory at the cutoff $\Lambda$.}
\hskip0.1cm 
The effective action of such a well-defined quantum field theory (\ref{bhlx}) at the cutoff $\Lambda$ and $G\gtrsim G_c$ is given by
\begin{eqnarray}
e^{iS^{\rm eff}}=Z\equiv \int_{\rm fields} e^{iL},
\label{eff}
\end{eqnarray}
where $L$ is the Lagrangian of quantum field theory (\ref{bhlx}) at the cutoff $\Lambda$.
Performing the Wick rotation $t\rightarrow \tau_{_E}=it$, $p^0\rightarrow p^0_{_E}=-ip^0$ and $p^2\rightarrow -p^2_{_E}$ to the Euclidean version of quantum field theory at the cutoff $\Lambda$, we have $Z\rightarrow Z_{_E}$, $e^{iS^{\rm eff}}\rightarrow e^{-S_{_E}^{\rm eff}}$ and  
\begin{eqnarray}
S_{_E}^{\rm eff} &=& - \int \frac{d^4x_{_E}d^4p_{_E}}{(2\pi)^4}{\rm Tr}\ln Z_{_E},
\label{effe}
\end{eqnarray}
where ${\rm Tr}$ indicates the sum over all the degree of freedom of the fields.
One can define the energy density of quantum field theories with the cutoff $\Lambda$ as
\begin{eqnarray}
{\mathcal E} &=& - \int_\Lambda \frac{d^4p_{_E}}{(2\pi)^4}{\rm Tr}\ln Z_{_E},
\label{effed_0}
\end{eqnarray}
which received all contributions from the bubble diagrams (closed loops) in the vacuum state. The energetic difference between the two configurations ($m_t=m_b\not=0$) and ($m_t\not=0, m_b=0$) is
\begin{eqnarray}
\Delta {\mathcal E} &=&{\mathcal E}(m_t=m_b\not=0)-{\mathcal E}(m_t\not=0, m_b=0).
\label{effed0}
\end{eqnarray}

Taking into account only the two-point Green functions (\ref{scalar}-\ref{fermion}) and (\ref{scalarb}-\ref{cpscalarb}), we approximately have 
\begin{eqnarray}
{\mathcal E} &\approx & + \int_\Lambda \frac{d^4p_{_E}}{(2\pi)^4}{\rm Tr}\sum_{\rm boson}\ln \Gamma^{-1}_{\rm boson}(p^2_{_E}) - \int_\Lambda \frac{d^4p_{_E}}{(2\pi)^4}{\rm Tr}\sum_{\rm fermion}\ln \Gamma^{-1}_{\rm fermion}(p^2_{_E}).
\label{effed}
\end{eqnarray}
The energetic difference (\ref{effed0}) between the two configurations ($m_t=m_b\not=0$) and ($m_t\not=0, m_b=0$) is approximately given by 
\begin{eqnarray}
\Delta {\mathcal E} 
&\approx & - 2N_c\int_\Lambda \frac{d^4p_{_E}}{(2\pi)^4}\ln\left( \frac{p^2_{_E}+m_b^2}{p^2_{_E}}\right)\nonumber\\
&+&2N_c\int_\Lambda \frac{d^4p_{_E}}{(2\pi)^4}\ln\left( \frac{\Gamma^{-1}_S(p^2_{_E}, m_{b})}{\Gamma^{-1}_S(p^2_{_E})}\right)\nonumber\\
&+&2N_c\int_\Lambda \frac{d^4p_{_E}}{(2\pi)^4}\ln\left( \frac{\Gamma^{-1}_P(p^2_{_E}, m_{b})}{\Gamma^{-1}_P(p^2_{_E})}\right)\nonumber\\
&+&8N_c\int_\Lambda \frac{d^4p_{_E}}{(2\pi)^4}\ln\left( \frac{\Gamma^{-1}_F(p^2_{_E}, m_{t,b})}{\Gamma^{-1}_F(p^2_{_E})}\right).
\label{effed1}
\end{eqnarray}
In this energetic difference $\Delta {\mathcal E}$, the contributions from the massive top quark, the intermediate gauge bosons ($W_\mu^\pm, Z_\mu^\circ$) and the massless photon are canceled. The first term is the difference between the massive and massless bottom-quark contributions. The second, third and fourth terms represent the vacuum-energy differences between the contributions of the composite scalar modes (\ref{scalar}-\ref{cpscalar}) in the $b$-channel for $m_b=m_t\not=0$ and the contributions of the inversed two-point Green functions in the $b$-channel for $m_b=0$.
Using Eqs.~(\ref{scalar}-\ref{cpscalarb}), we calculate the energetic difference (\ref{effed1}) and obtain the leading order:
\begin{eqnarray}
\Delta {\mathcal E} 
&\approx & +\frac{N_c}{\pi^2}\,\left[
0.18\,\Lambda^4\left(\frac{G_c}{G}\right)-0.36\,(\Lambda m_t)^2\ln\left(\frac{\Lambda}{m_t}\right)^2 
%+0.08\,(\Lambda m_t)^2
\right] > 0,
\label{result}
\end{eqnarray}
which shows that the configuration ($m_t\not=0, m_b=0$) is energetically 
more favorable than the configuration ($m_t=m_b\not=0$). By using Eqs.~(\ref{scalarbb}-\ref{delta1}), we obtain the analog of
Eq.~(\ref{result}) in the case that the top-quark mass $m_t$ is generated by the spontaneous chiral symmetry breaking and that the bottom-quark mass $\bar m_b$ is induced by an explicit chiral symmetry breaking:
\begin{eqnarray}
\Delta {\mathcal E} 
&\approx & 
+\frac{N_c}{\pi^2}\left[0.54\times 4\pi^2  \delta-0.36(\Lambda^2m^2_t)\ln\left(\frac{\Lambda}{m_t}\right) + 0.18(\Lambda^2\bar m^2_b)\ln\left(\frac{\Lambda}{\bar m_b}\right)\right]\nonumber\\
&= &
+0.09\,\frac{N_c}{\pi^2}\left[(\Lambda^2m^2_t)\ln\left(\frac{\Lambda}{m_t}\right)^2 + (\Lambda^2\bar m^2_b)\ln\left(\frac{\Lambda}{\bar m_b}\right)^2\right] >0,
\label{result2}
\end{eqnarray}
where $m_b\ll m_t$.

These results (\ref{result}) and (\ref{result2}) are not surprising. One can see that the vacuum energy decreases (the system of fields gains energy) as the fermions acquire their masses by the spontaneous chiral symmetry breaking; 
the first line of Eq.~(\ref{effed1}), however, the associated scalar and pseudoscalar modes, the second, third and fourth lines of Eq.~(\ref{effed1}), have positive contributions to the vacuum energy. 
Three pseudoscalar (Goldstone) modes become the longitudinal modes of the intermediate gauge bosons $W_\mu^\pm$ and $Z^\circ$. As more fermions acquire their masses by the spontaneous chiral symmetry breaking, more associated scalar and pseudoscalar modes are produced. As a result, the energetically favorable configuration is the one in which only one quark (the top quark) acquires its mass by the spontaneous chiral symmetry breaking, with three pseudoscalar modes as the longitudinal modes of the massive gauge bosons and a scalar particle of the Higgs type. 

These discussions and calculations can be generalized to the situation involves all three quark families: 
\begin{eqnarray}
L 
&=& L_{\rm kinetic} + G(\bar\Psi^{ia}_L\psi^{(+)j}_{Ra})(\bar \psi^{(+)b}_{jR}\Psi_{Lib})
+ G(\bar\Psi^{ia}_L\psi^{(-)j}_{Ra})(\bar \psi^{(-)b}_{jR}\Psi_{Lib}),
\label{bhlx_e}
\end{eqnarray}
where $\Psi_{Lib}$ represents the left-handed doublet and the index $i,j=1,2,3$ represents three quark families; the right-handed singlet $\psi^{(\pm)j}_{Ra}$ represents the charged $+2/3$ quarks and the charge $-1/3$ quarks. In addition to the symmetries in Eq.~(\ref{bhlx}), the Lagrangian (\ref{bhlx_e}) has the global flavor symmetry of quark families $SU_L(3)\times U^{(\pm)}_R(3)$. This result indicates that only one massive quark acquires its mass via the spontaneous chiral-symmetry breaking from the energetic point of view, whereas the other quarks must acquire their masses via an explicit chiral symmetry breaking without additional scalar and pusdoscalar modes. In conclusion, the configuration in which only one quark (the top quark) acquires its mass via the spontaneously dynamical symmetry breaking is an energetically favorable configuration (the ground state) of the quantum field theory with the high-dimension operators of all the fermion fields, which originated from new dynamics at the cutoff $\Lambda\gg m_t$.

It is then interesting to ask the question of how leptons acquire their masses. Provided that all symmetries of the Standard Model are respected, there are no principles and prerequisites to preclude one from generalizing Eq.~(\ref{bhlx}) to the lepton sector with the same coupling $G$ at the cutoff $\Lambda$. From the gap equation (\ref{gap0}), we find that $2GN_c$ is an effective coupling for the spontaneous generation of the top quark mass if the $G$-value is larger (near) than (to) a critical value. However, the effective coupling for the gap equation of the colorless lepton sector 
is $2G$, which is smaller than $2GN_c$. This result implies that the gap equation for the lepton sector cannot be satisfied for a nontrivial solution $m\not=0$, with the same $G$-value for a nontrivial solution $m_t\not=0$ to the gap equation of the top quark, the gap equation of the lepton sector only admits the trivial solution $m=0$. Leptons must acquire their masses via an explicit chiral symmetry breaking. These discussions and calculations should be properly performed in the framework of unification theories of the lepton and quark sectors, for instance, the $SU(5)$ theory at the cutoff $\Lambda$.            
The relations of the gauge couplings, the coupling $G$ at the cutoff $\Lambda$ and the mass scales of the top quark and composite scalar particle were discussed and calculated in the framework of renormalization-group equations \cite{bhl1990,Kahana}. In future works, we will draw our attention to these relations within the following contexts: (i) an explicit chiral symmetry breaking mechanism and (ii) the parity-conserving fermion spectrum and gauge couplings at the cutoff $\Lambda$ that are briefly discussed below. 
    
%\vskip0.1cm
\noindent
{\bf Explicit chiral symmetry breaking for fermion masses.}
\hskip0.1cm 
To end our Letter, we present, on the basis of our previous works on this issue, a brief discussion on (i) high-dimensional operators due to the quantum gravity at the Planck length 
($a_{\rm pl}\sim 10^{-33}\,$cm, $\Lambda_{\rm pl}=\pi/a_{\rm pl}\sim 10^{19}\,$GeV) and (ii) the parity-conserving fermion spectrum and gauge couplings at the cutoff $\Lambda$, leading to an explicit chiral symmetry breaking mechanism for the mass generation of fermions except the top quark. 

We may conceive that the physical space-time is endowed with the Planck length $a_{\rm pl}$, which precisely due to the violent fluctuations that the gravitational field {\it must}
exhibit at $a_{\rm pl}$, and thus the basic arena of physical reality becomes a {\it random lattice with a lattice constant} $a\sim a_{\rm pl}$ \cite{preparata91}. We recently calculated this minimal length $a\approx 1.2\,a_{\rm pl}$ \cite{xue2010} while
studying the quantum Einstein-Cartan theory in the framework of Regge calculus and its variant \cite{wheeler1964,regge61,hammer_book}. 

This discrete space-time provides a natural regulator for local quantum field theories of particles and gauge interactions. A natural regularized quantum field theory demands the existence of ultra violet fixed points where the renormalization group invariance follows so that the low-energy observables are independent of high-energy cutoff. Based on low-energy observations of parity violation, the Lagrangian of Standard Model was built in such a way as preserve the exact chiral gauge symmetries $SU_L(2)\otimes U_Y(1)$ that are accommodated by left-handed fermion doublets and right-handed fermion singles. However, a profound result, in the form of a generic no-go theorem \cite{nn1981,nn1991}, tells us that there is no consistent way to straightforwardly transpose on a discrete space-time the bilinear fermion Lagrangian of the continuum theory in such a way as to preserve the chiral gauge symmetries exactly. We were led to consider at least quadrilinear (four) fermion interactions to preserve the chiral gauge symmetries \cite{xue_chiral}. The very-small-scale structure of space-time and high-dimensional operators of fermion interactions must be very complex \cite{torsion}. In addition to the 6-dimensional interaction of Eq.~(\ref{bhlx_e}), we introduced a 10-dimensional interaction of four fermion fields \cite{xue_chiral}:
\begin{eqnarray}
G_2(\bar\Psi^{ia}_L\Delta\psi^{(+)j}_{Ra})(\bar \Delta\psi^{(+)b}_{jR}\Psi_{Lib})
+ G_2(\bar\Psi^{ia}_L\Delta\psi^{(-)j}_{Ra})(\Delta\bar \psi^{(-)b}_{jR}\Psi_{Lib}),
\label{bhlx_e1}
\end{eqnarray}
where $\Delta\psi^{(\pm)j}_{Ra}=\psi^{(\pm)j}_{Ra}(x+a)+\psi^{(\pm)j}_{Ra}(x-a)-2\psi^{(\pm)j}_{Ra}(x)$ and the strong coupling $a^2G_2\gg 1$. The Ward identity of the shift symmetry $\psi_R(x)\rightarrow \psi_R(x) + {\rm const.}$ leads to the one-particle irreducible vertex of the four-fermion interaction, 
\begin{eqnarray}
\Gamma^{(4)}(p-p')=a^2G_2\,w(p+q/2)\,w(p'+q/2),\quad  w(k)\equiv 2\sin^2(k\,a/2)
\label{4fe}
\end{eqnarray}
where $p+q/2$ and $p'+q/2$ are the momenta of the $\psi_R$ field and $p-q/2$ and $p'-q/2$ are the momenta of the $\Psi_L^i$ field ($q$ is the momentum transfer). 

At high energies, namely, at large fermion energy transfer, $\Gamma^{(4)}$ is so large that bound states of three fermions,
\begin{eqnarray}
\Psi^i_R\sim (\bar\psi_R\Psi^i_L)\psi_R
\label{3f}
\end{eqnarray}
are formed and carry a chirality that is opposite that of left-handed fermions $\Psi^i_L$. Namely, $\Psi^i_R$ is a right-handed $SU_L(2)$ doublet.
In addition, $\Psi^i_L$ combines with $\Psi^i_R$ to form a Dirac fermion $\Psi^i_D=(\Psi^i_L, \Psi^i_R)$, whose inverse propagator is obtained by the method of strong coupling expansion in terms of $1/(a^2G_2)$ 
\cite{xue_chiral, xue2008}:
\begin{eqnarray}
S^{-1}_{ij}(k)= \delta_{ij}\frac{i}{a}\gamma_\mu 
\sin(k_\mu a)+ \delta_{ij}M(k),
\label{infp}
\end{eqnarray}
where 
$M(k)=aG_2\,w(k)$ is the chiral-gauge-symmetric masses of composite Dirac fermions \cite{neutral}.
Left-handed fermions $\Psi^i_L$ and right-handed composite fermions $\Psi^i_R$ couple to intermediate gauge bosons in the same way. 
Therefore the effective one-particle irreducible interacting vertex of fermions ($p,p'$) and $W^\pm$ gauge boson ($q=p'-p$)  \cite{xue_parity, r-neutrino} is the following: 
\begin{eqnarray}
\Gamma^{ij}_\mu (p,p')&=&i\frac{g_2}{2\sqrt{2}}U_{ij}\gamma_\mu \left[P_L+f(p,p')\right]
\label{wv}
\end{eqnarray}
where $g_2$ is the $SU_L(2)$ coupling, $U_{ij}$ the CKM matrix, and the
left-handed projector $P_L=(1-\gamma_5)/2$. The vector-like (parity conserving) form factor $f(p,p')$ is related to the chiral-symmetric mass $M(k)$ in Eq.~(\ref{infp}) by the Ward identity of chiral gauge symmetries. At high energies, $f(p,p')\not=0$ %for $p,p'\ge{\mathcal E}$, 
and chiral-gauge coupling becomes vector-like. Consequently the parity symmetry is conserved \cite{xue_chiral,xue_parity,x2011}.

At low energies, namely, at small fermion energy-momenta $(p,p')$ and energy transfer $(q)$, the interacting vertex $\Gamma^{(4)}(p,p')$ of Eq.~(\ref{wv}) becomes so small that the binding energies $E_{\rm bind}[G_2(a),a]$ of the bound states (\ref{3f}) of three fermions 
vanish. As a result, these bound states (\ref{3f}) dissolve into their constituents \cite{pole-cut}, and the mass term $M(k)$ and the vector-like form factor $f(p,p')$ vanish. This result restores the chiral-gauged fermion spectra and couplings, as described by the Standard Model \cite{c-symmetry}. This action is what we called the dissolving phenomenon of the bound states of three fermions at the energy threshold ${\mathcal E}_{\rm thre}$.
However, it is complicated to quantitatively show this dissolving phenomenon and calculate the scale of energy threshold ${\mathcal E}_{\rm thre}$. Non-perturbative calculations are required to find a ultra violet fixed point and a renormalization group equation for $G_2(a,{\mathcal E}_{\rm thre})$ in the neighborhood of the ultra violet fixed point, where the 
high-dimensional operator (\ref{bhlx_e1}) receives anomalous dimensions and becomes renormalizable 4-dimensional operators.    
Nevertheless, we postulated \cite{xue_chiral,xue_parity} that the  energy threshold ${\mathcal E}_{\rm thre}$ coincides to the cutoff $\Lambda$ at which the four-fermion operators (\ref{bhlx_e}) are relevant operators and
\begin{equation}
\Lambda_{\rm EW}\ll \Lambda < \Lambda_{\rm pl},\quad \Lambda={\mathcal E}_{\rm thre},
\label{epsilon}
\end{equation}
where $\Lambda_{\rm EW}\sim 250\,$GeV is the electroweak-symmetry-breaking scale. 

Due to the vector-like form factor in the interacting vertex (\ref{wv}) for high energies, the Dyson equations for
the self-energy functions of the quarks ($q=2/3$) and the quarks ($q=-1/3$) are coupled. If the top quark acquires its mass by the spontaneous symmetry breaking, the top-quark mass term enters the Dyson equations for other fermion self-energy functions as an explicit chiral-symmetry-breaking term; as a result, the bottom quark acquires its mass \cite{gxtb,xfine}, and the other quarks get their masses from the mixing matrix $U_{ij}$ \cite{x1997}. If the quarks are massive, the Dyson equations for
the lepton self-energy functions receive explicit breaking terms via the vector-like interacting vertex (\ref{wv}) with $W_\mu^\pm$. As a result, the Dyson equations for
the lepton self-energy functions have nontrivial solutions for massive leptons \cite{gx1993}. Beside the mixing matrix $U_{ij}$ of the lepton sector also introduces explicit symmetry breaking terms in the Dyson equations for lepton self-energy functions \cite{x1999}.

In Ref.~\cite{xfine}, we
solved the coupled gap equations (Dyson equations) for the top and bottom quarks' self-energy functions simultaneously.
As a result, we found that the ratio of the top
and bottom masses is given by $(m_b/m_t)^2\simeq \alpha/(3\pi)$, and the unnatural fine-tuning problem is overcome, provided
that the quadratic divergence is removed by setting the four-fermion coupling
$G/G_c=1+{\mathcal O}(m_b^2/m_t^2)$ instead of drastically fine-tunning the four-fermion coupling (\ref{delta}), $G/G_c=1+ {\mathcal O}(m_t^2/\Lambda^2)$ for $\Lambda \gg m_t$. In this case, without drastic fine-tuning, one can have the physically sensible formula that connects the pseudoscalar (channels (\ref{pscalarb},\ref{cpscalarb}) coupling to the longitudinal $W$ and $Z$) decay constant $f_\pi$ to the top-quark mass \cite{bhl1990}:
\begin{equation}
f_\pi^2=\frac{1}{4\sqrt{2}G_F}\approx \frac{N_c}{32\pi^2}m_t^2\ln \frac{\Lambda^2}{m_t^2},
\label{decay}
\end{equation}
for $\Lambda\sim 10^{15}$GeV and $m_t\approx 173$GeV, where $G_F$ is the Fermi constant. This is regarded as a general discussion, and we should calculate the pesudoscalar decay constant $f_\pi$ by duly taking into account the coupled gap equations (Dyson equations) for the top and bottom quarks due to the vector-like gauge coupling (\ref{wv}) above the energy threshold ${\mathcal E}_{\rm thre}$ of Eq.~(\ref{epsilon}). In this framework, it is also necessary to study the scalar channel (\ref{scalarb}) to see the mass of the composite particle of the Higgs type. We must confess that, so far, we have not yet obtained a massive composite scalar particle in the connection with the observed 125 GeV particle of the Higgs type. To obtain the elementary fermion and composite scalar spectra, we must solve coupled Dyson equations for fermions and composite scalars due to the vector-like coupling (\ref{wv}) at the energy threshold (\ref{epsilon}) in the context of renormalization-group equations evolving from the scale $\Lambda$ to the low-energy attractive fixed point.
We will attempt to perform these difficult tasks in future.

%\vskip0.1cm
%\noindent
%{\bf Acknowledgment.}
%\hskip0.1cm


\begin{thebibliography}{99}
\bibitem{r-neutrino}
Similar discussions and calculations for chiral-gauge coupling to the neutral gauge boson $Z^0$ can also be found in Ref.~\cite{xue_chiral, xue_parity}.
\bibitem{pole-cut}
Bound states of three fermions are presented as poles of four-fermion interacting $S$-matrix, locating at $E_{\rm bind}[G_2(a),a]<0$ in the energy-momentum plane. When the coupling $a^2G_2(a)$ becomes smaller, $E_{\rm bind}\rightarrow 0^-$, these poles of the four-fermion interacting $S$-matrix become cuts, representing three elementary fermion states.
\bibitem{c-symmetry} In the phase transition of fermion bound states $\Psi_R\sim (\bar\psi_R\Psi_L)\psi_R$ formation (dissolution) from (into) their constitutes, chiral symmetries are exactly preserved, and realized by different chiral fermion contents \cite{xue_chiral}.  It
should be pointed out that this phenomenon is not a spontaneous symmetry
breaking and no Goldstone bosons are produced \cite{njl}.
\end{thebibliography}
\end{document}